\documentclass[aps,reprint,floatfix,superscriptaddress]{revtex4-2}
\usepackage[utf8]{inputenc}
\usepackage{amsmath}
\usepackage{amssymb}
\usepackage{graphicx}
\usepackage{xcolor}
\usepackage{hyperref}
\usepackage[normalem]{ulem}
\usepackage{siunitx}            
\usepackage[version=4]{mhchem}  
\usepackage{physics}            
\graphicspath{{figures/}}

\let\vec\oldvec
\newcommand{\vec}[1]{\ensuremath{\boldsymbol{#1}}}
\newcommand{\qmop}[1]{\ensuremath{\hat{#1}}}

\DeclareMathOperator{\sgn}{sgn}

\newcommand{\ie}{{\it i.e.~}} 	

\begin{document}
\title{Supercurrent interference in HgTe Josephson junctions}

\author{Wolfgang Himmler}
\altaffiliation{These authors contributed equally to this work.}
\affiliation{Institute of Experimental and Applied Physics, University of Regensburg, D-93040 Regensburg, Germany}
\email{corresponding authors: wolfgang.himmler@ur.de,\\ dieter.weiss@ur.de}
\author{Ralf Fischer}
\altaffiliation{These authors contributed equally to this work.}
\affiliation{Institute of Experimental and Applied Physics, University of Regensburg, D-93040 Regensburg, Germany}
\author{Michael Barth}
\affiliation{Institute of Theoretical Physics, University of Regensburg, D-93040 Regensburg, Germany}
\author{Jacob Fuchs}
\affiliation{Institute of Theoretical Physics, University of Regensburg, D-93040 Regensburg, Germany}
\author{Dmitriy A. Kozlov}
\affiliation{Institute of Experimental and Applied Physics, University of Regensburg, D-93040 Regensburg, Germany}
\affiliation{A. V. Rzhanov Institute of Semiconductor Physics, Novosibirsk 630090, Russia}
\author{Nikolay N. Mikhailov}
\affiliation{A. V. Rzhanov Institute of Semiconductor Physics, Novosibirsk 630090, Russia}
\author{Sergey A. Dvoretsky}
\affiliation{A. V. Rzhanov Institute of Semiconductor Physics, Novosibirsk 630090, Russia}
\author{Christoph Strunk}
\affiliation{Institute of Experimental and Applied Physics, University of Regensburg, D-93040 Regensburg, Germany}
\author{Cosimo Gorini}
\affiliation{Universit\'{e} Paris-Saclay, CEA, CNRS, SPEC, 91191 Gif-sur-Yvette, France}
\author{Klaus Richter}
\affiliation{Institute of Theoretical Physics, University of Regensburg, D-93040 Regensburg, Germany}
\author{Dieter Weiss}
\affiliation{Institute of Experimental and Applied Physics, University of Regensburg, D-93040 Regensburg, Germany}

\date{\today}

\begin{abstract}
Wires made of topological insulators (TI) are a promising platform for searching for Majorana bound states. These states can be probed by analyzing the fractional ac Josephson effect in Josephson junctions with the
TI wire as a weak link. An axial magnetic field can be used to tune the system from  trivial to topologically nontrivial. Here we investigate the oscillations of the supercurrent in such wire Josephson junctions as a function of the axial magnetic field strength and different contact transparencies. Although the current flows on average parallel to the magnetic field we observe $h/2e$, $h/4e$- and even $h/8e$-periodic oscillations of the supercurrent in samples with lower contact transparencies.
Corresponding tight-binding transport simulations using a Bogoliubov-de Gennes model Hamiltonian yield the supercurrent through the Josephson junctions, showing in particular the peculiar $h/4e$-periodic oscillations observed in experiments. A further semiclassical analysis based on Andreev-reflected trajectories connecting the two superconductors
allows us to identify the physical origin of these oscillations.
They can be related to flux-enclosing paths winding around
the TI-nanowire, thereby highlighting the three-dimensional character
of the junction geometry compared to common planar junctions.
\end{abstract}

\maketitle

\section{Introduction}

In wires made from a three-dimensional topological insulator (3DTI) the topological surface states form a two-dimensional conducting electron layer that envelops the bulk. The energy spectrum of these wires features a gap at zero magnetic field which closes when an axial magnetic flux of $\phi_0 = h/2e$ threads the wire cross section \cite{ostrovsky2010,Bardarson-et-al-2010,Bardarson_2013,Kozlovsky_2020,graf2020}. With closing of the gap, a  non-degenerate perfectly transmitting mode appears, rendering the wire’s band structure topologically nontrivial. Whereas semiconductor wires with strong spin-orbit interaction are the so far prevailing material platform to search for Majorana bound states (\cite{Mourik2012,Rokhinson2012,Deng2016,Gul2018} and references therein), mesoscopic wires made of TI material are a promising alternative \cite{Cook-Vazifeh-Franz_2012,Ilan-et-al-2014,deJuan-et-al-2014,deJuan-et-al-2019,Xypakis-et-al-2020,Legg-et-al-2021,Legg_2022}. Recent experiments utilizing the fractional Josephson effect in \ce{HgTe}-based Josephson junctions (JJ) indeed  provided evidence that the 4$\pi$-periodic supercurrents observed for an axial magnetic flux $>\phi_0/4$ are of topological origin \cite{Fischer-et-al-2022}. In these experiments two superconducting contacts are placed across a HgTe wire with the TI wire constituting the normal region forming a JJ. 

So far, semiconductor wires with strong spin-orbit interaction have been the prevailing system class to search for topological superconductivity. The JJ built from such wire structures and their behavior in a magnetic field have been investigated, for instance, in Refs.~\cite{Suominen-et-al-2017,Zuo-et-al-2017,Gharavi_2017,2019-PhysRevB-Sriram_et_al,Kringhoj_2021,Perla_2021,Stampfer_2022}. Related work is also available on JJs made from \ce{Cd3As2} \cite{Li-et-al-2021} or \ce{Bi2Se3} \cite{Chen-et-al-2018} wires and \ce{(Bi,Sb)2Te3} nanoribbons \cite{Schueffelgen2019}.

Here we investigate the evolution of  supercurrent interference in \ce{HgTe} wire-based JJs as a function of an axial magnetic field. The supercurrent flows between the two superconducting (sc) contacts along the TI and is driven by the difference $\varphi$ of the superconducting phase between the two contacts. In the presence of an axial magnetic field the supercurrent amplitude oscillates as a result of interference between Andreev bound states acquiring different phases along their quasi-classical trajectory between the sc contacts \cite{2016-PhysRevB-Ostroukh_et_al}. In the case of a  junction with the magnetic field perpendicular to the supercurrent, the supercurrent oscillations are described by $I_c(\phi)=I_c(0)\abs{\sin(\pi\phi/\phi_0)/(\pi\phi/\phi_0)}$,    
identical to the Fraunhofer pattern of a single slit experiment. Such a pattern is not expected for the TI-wire JJ mentioned above, since the current flows on average parallel to the magnetic field and the shortest ballistic trajectories should not pick up any phase from the magnetic flux. Remarkably, we find oscillations in the supercurrent which are $h/2e$, $h/4e$ and even $h/8e$ periodic, thus constituting a highly unusual interference pattern. Below we relate these findings, both experimentally and theoretically in a consistent way, to the three-dimensional JJ-geometry and the coupling of the superconducting contacts to the TI wire.

\section{Device parameters and experimental setup}

\begin{figure*}
    \centering
    \includegraphics{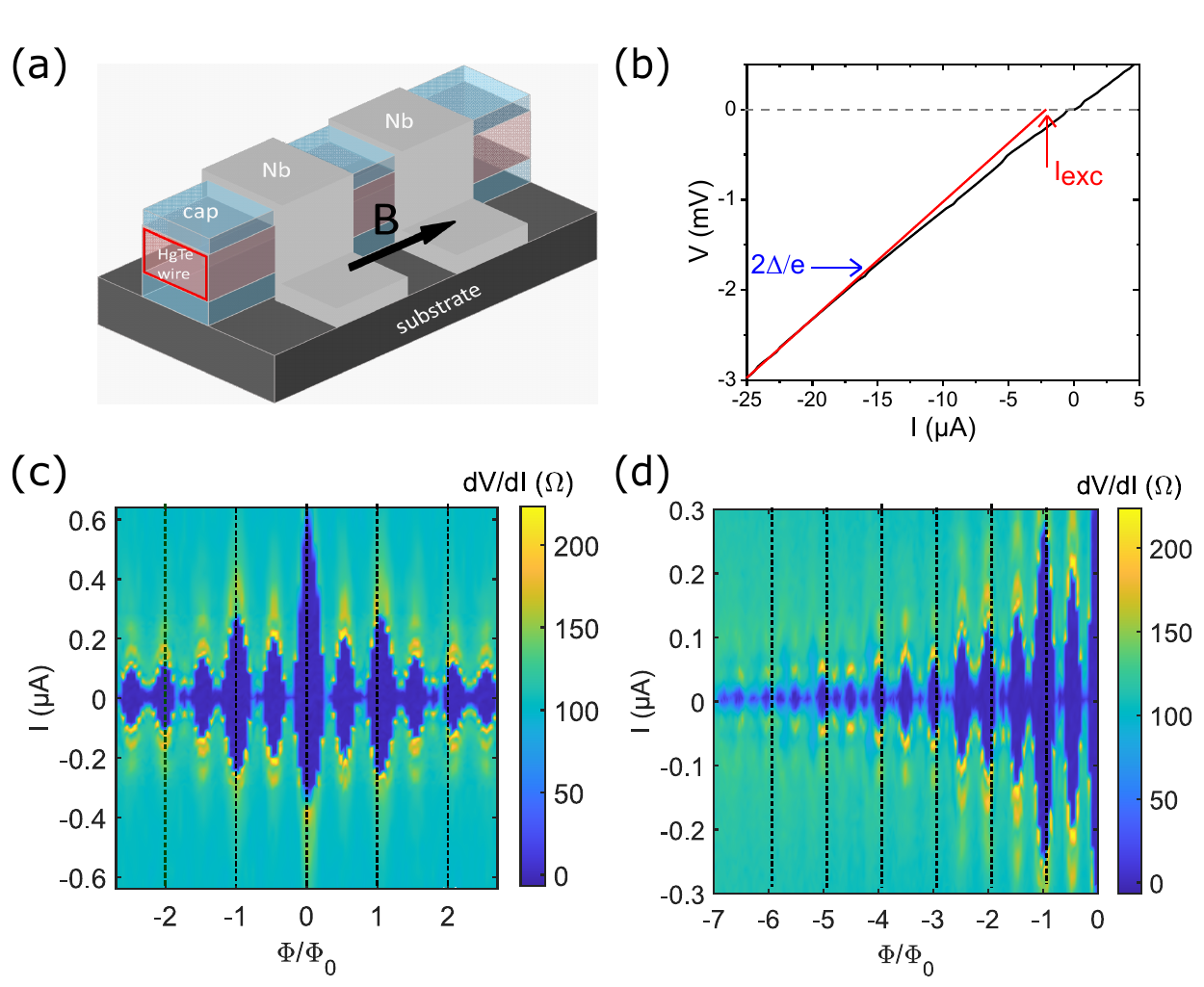}
    \caption{\textbf{Sample layout, excess current, and critical current vs. magnetic field $\boldsymbol{B}$.} \textbf{a,} Cartoon of the sample layout showing the \ce{HgTe} wire and the two \ce{Nb} contacts which form the Josephson junction. The topological surface states are shown in red. The magnetic field is oriented parallel to the axis of the wires. \textbf{b,} $I$-$V$ trace of nanowire G (black trace) for $B=0$ at a temperature of $\SI{27}{mK}$. For high bias voltages, the slope represents the normal-state resistance $R_N$, while for lower voltages Andreev reflections influence the trace resulting in an excess current $I_{exc}=\SI{2.2}{\micro \ampere}$. The superconducting gap $\Delta$ can be extracted from the curve as the trace starts to deviate from the constant normal-state resistance (red curve) if $eV<2\Delta \approx \SI{1.8}{mV}$. With these values the parameter $Z\approx0.98$\ is estimated. Thus, the transparency is given by $D\approx0.51$. \textbf{c,} Color map of the differential resistance $dV/dI$ of sample G as a function of the current $I$ and the magnetic flux $\phi/\phi_0$ ($\phi_0=h/2e$). For sample G, $\phi/\phi_0$ corresponds to $B\approx\SI{36}{mT}$. Superconducting regions are shown in blue. The critical current oscillates with a period $\phi_0/2$, while the side maxima at $\phi=\phi_0$ are most pronounced. \textbf{d,} Color map of the differential resistance $dV/dI$ of sample G up to higher values of the magnetic flux. For $\phi/\phi_0>3$, additional maxima appear resulting in a $\phi_0/4$ periodicity.}
    \label{fig:device_excess_sampleG}
\end{figure*}


We considered 9 devices (labeled A-J, ordered by descending JJ transparency D) made from wafers with an \SI{80}{nm} thick, strained \ce{HgTe} film, which is grown on \ce{CdTe} by molecular beam epitaxy. A thin \ce{Cd_{0.7}Hg_{0.3}Te} buffer layer was introduced in between to improve the quality of the samples \cite{2014-PhysRevLett-Kozlov_et_al}. Finally, the wafers are capped by \ce{Cd_{0.7}Hg_{0.3}Te} and \ce{CdTe}. Fig.~\ref{fig:device_excess_sampleG}a 
sketches the wafer structure and the device. Typically, the Fermi level $\mu$ is located at the top of the valence band, and surface electrons as well as bulk holes co-exist. The electron density is of order $n_e\sim\SI{e11}{\metre^{-2}}$. Additionally, \ce{In}-doping was added to the \ce{Cd_{0.7}Hg_{0.3}Te} layers for specific wafers (sample D, G). This increases the 
electron density up to one order of magnitude, 
since the Fermi level $\mu$ is shifted to the conduction band.
We fabricate the nanowires using electron beam lithography and wet-chemical etching \cite{2018-PhysRevB-Ziegler_et_al, Fischer-et-al-2022}. Due to the wet-chemical etching, the wires have a trapezoidal cross-section. In the following, we use the average width, which typically ranges between $\SIrange{500}{700}{\nano\metre}$. The wire perimeter is always shorter than the phase coherence length, which is of the order of several microns \cite{2018-PhysRevB-Ziegler_et_al}, and transport is thus coherent. The superconducting \ce{Nb}  contacts are placed on the surface of \ce{HgTe} after removing the capping layers by wet-chemical etching. To enhance contact quality, we clean the \ce{HgTe} surface by gentle in-situ \ce{Ar}$^+$-sputtering and add a thin \ce{Ti} layer ($\sim\SI{3}{nm}$), grown in-situ by thermal evaporation, below the \ce{Nb}. As \ce{Nb} tends to oxidize, we add a thin layer of \ce{Pt} to protect the \ce{Nb}. The distance between adjacent superconducting contacts is between $\SIrange{50}{240}{\nano\metre}$. 
The samples are cooled down in a dilution refrigerator with a base temperature of $\SI{27}{mK}$. The $B$-field is aligned parallel to  the wire's axis so that the magnetic flux through the wire is $\phi=BA$, where $A$ is the cross-sectional area of the wire. The measurements are taken using standard dc techniques, while the dc lines are filtered by $\pi$-filters at room temperature and \ce{Ag}-epoxy filters \cite{2014-APL-Scheller_et_al} as well as RC-filters in the mixing chamber. The differential resistance $dV/dI$ is measured by superimposing the dc bias with a small ac signal using lock-in amplifiers.

The transparency of the superconducting contacts is determined by voltage-biased measurements. An $I$-$V$ trace, exemplary shown for sample G, is plotted in Fig.~\ref{fig:device_excess_sampleG}b. The slope of the trace stays constant and represents the normal-state resistance for bias voltages $V > \SI{1.8}{mV}$, while for lower voltages Andreev reflections modify the slope \cite{1982-PhysRevB-Blonder_Tinkham_Klapwijk, 1983-PhysRevB-Octavio_et_al}. The change of the resistance gives an estimation for the superconducting gap of \ce{Nb} $\Delta = eV/2\approx \SI{0.95}{meV}$. The additional current flowing across the junction is the excess current $I_{exc}=\SI{2.2}{\micro \ampere}$. 
With the extracted values, we estimate the dimensionless parameter $Z$ which describes the average transparency $D=1/(1+Z^2)$ using the expression of Niebler et al. \cite{2009-SupScience-Niebler_et_al}, which is based on the work of Flensberg et al. \cite{1988-PhysRevB-Flensberg_et_al} and the OBTK-theory \cite{1983-PhysRevB-Octavio_et_al}.  Inserting the values of sample G, we get $Z \approx0.98$ and $D\approx0.51$ \footnote{We are aware that OBTK-theory fails for low junction transparencies, because it does not correctly take into account interferences from multiple reflections (see, e.g., [13-15] in the supplement to \cite{Fischer-et-al-2022}}.
An overview of the individual sample geometries and transparencies is displayed in table \ref{tab:sample-table}.

\begin{table}[]
\centering
\begin{tabular}{l|c c c c c c c c c}
Sample                & A    & B    & C    & D    & E    & F    & G    & H    & J    \\
\hline
$D$~        & 0.70 & 0.66 & 0.64 & 0.63 & 0.62 & 0.57 & 0.51 & 0.49 & 0.43 \\
$w$ $[\si{\nano\metre}]$~       & 450  & 900  & 570  & 600 & 470  & 700  & 700 & 540  & 520    \\
$L$ $[\si{\nano\metre}]$~      & 50   & 160  & 100  & 180 & 65   & 70   & 110  & 40   & 240    \\
$W_{\text{S}}$ $[\si{\micro\metre}]$~ & 1.3  & 1.3  & 1.3  & 1.3  & 2.3  & 4.3  & 0.6  & 4.3  & 0.6  
\end{tabular}

\caption{\textbf{Sample transparencies and geometries.} Junction width $w$, length $L$ and width $W_S$ of the deposited superconducting niobium fingers for samples A-J, ordered by sample transparency D.}
\label{tab:sample-table}

\end{table}

\begin{figure*}
    \centering
    \includegraphics{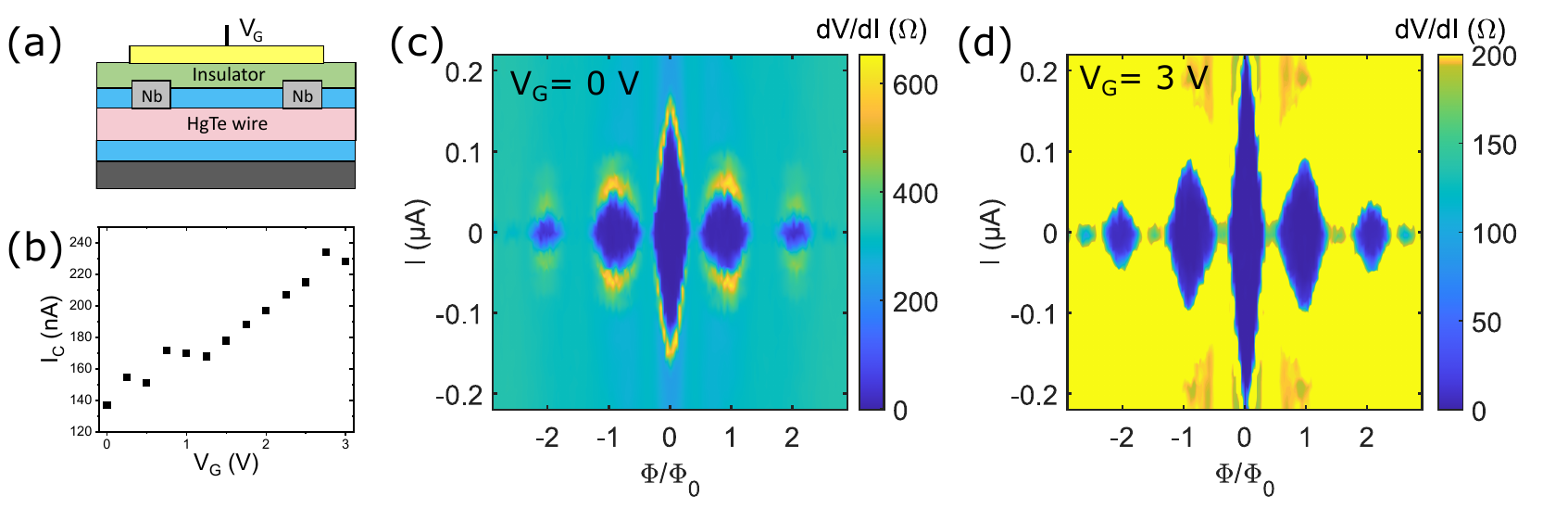}
    \caption{\textbf{Gate dependence of $\boldsymbol{I_C(B)}$-oscillations. } \textbf{a,} Sketch of the sample layout. An insulator and a metallic topgate is placed on top of the junction. \textbf{b,} The critical current $I_C$ increases for higher gate voltages $V_G$. \textbf{c,} Color map of the differential resistance $dV/dI$ of sample J as a function of the current I  and the magnetic flux $\phi/\phi_0$ at $V_G=0$. The critical current oscillates with a period of $\phi_0$. \textbf{d,} The corresponding color maps at $V_G=\SI{3}{V}$. Additional oscillations of $I_C$ appear recovering the $\phi_0/2$ periodicity.}
    \label{fig:gated device}
\end{figure*}


\section{Experimental Results}
\label{sec:exp}

In a Josephson junction, a magnetic field parallel to the current direction is expected to act as a  pair-breaker \cite{Yip2000,Heikkila2000,Crosser2008}. In this scenario, the critical current of the device  decreases monotonously with increasing magnetic field strength. For some  of our devices, however, we found a strong modulation of the critical current $I_C$ as a function of the axial magnetic field $B$. Fig.~\ref{fig:device_excess_sampleG}c presents a color map of the differential resistance $dV/dI$ for sample G as a function of  current $I$ and  magnetic flux $\phi$, threading the cross-sectional area of the nanowire. This device has a critical current $I_C\approx\SI{600}{\nano\ampere}$ and shows the most prominent oscillations of $I_C$. With the width of the wire $w\approx\SI{700}{\nano\metre}$, one superconducting flux quantum $\phi_0=h/2e$ corresponds to $B\approx\SI{36}{mT}$.  Blue regions in the color map illustrate superconducting states. The pattern displays maxima of $I_C$ for $\phi=n\cdot\phi_0/2$ with  $n$ an integer, while $I_C$ is fully suppressed in between them. Furthermore, the maxima at multiples of $\phi_0$ are more pronounced than the $\phi_0/2$ maxima. Data of the same device up to higher fluxes is shown in Fig.~\ref{fig:device_excess_sampleG}d. Here, additional maxima at $\phi=n\cdot\phi_0/4, n\in\mathbb{Z}$, appear. The $h/4e$ periodicity eventually changes to $h/8e$ at higher magnetic fields. The envelope of this pattern can be ascribed to the expected pair-breaking mechanism.  We note at this point that roughly $h/2e$ periodic oscillations were observed by Stampfer et al. and ascribed to oscillations of the transmission due to the conventional Aharonov-Bohm effect \cite{Stampfer_2022}. Nonmonotonic behavior of $I_C(B)$ with multiple nodes and lobes but without clear periodicity were observed in semiconductor nanowire JJs in an axial field \cite{Gharavi_2017,Zuo-et-al-2017}.

Only a fraction of the investigated junctions show an oscillatory interference pattern of the critical current as a function of the flux, while the critical current monotonously decreases with the magnetic field for other samples. Even the exact shape and periodicity of the pattern, if it exists, differs for various devices. Therefore, we will analyze the emergence of the $I_C(B)$-oscillations in the following for different experimental parameters like contact transparency and gate voltage.

\begin{figure*}
    \centering
    \includegraphics{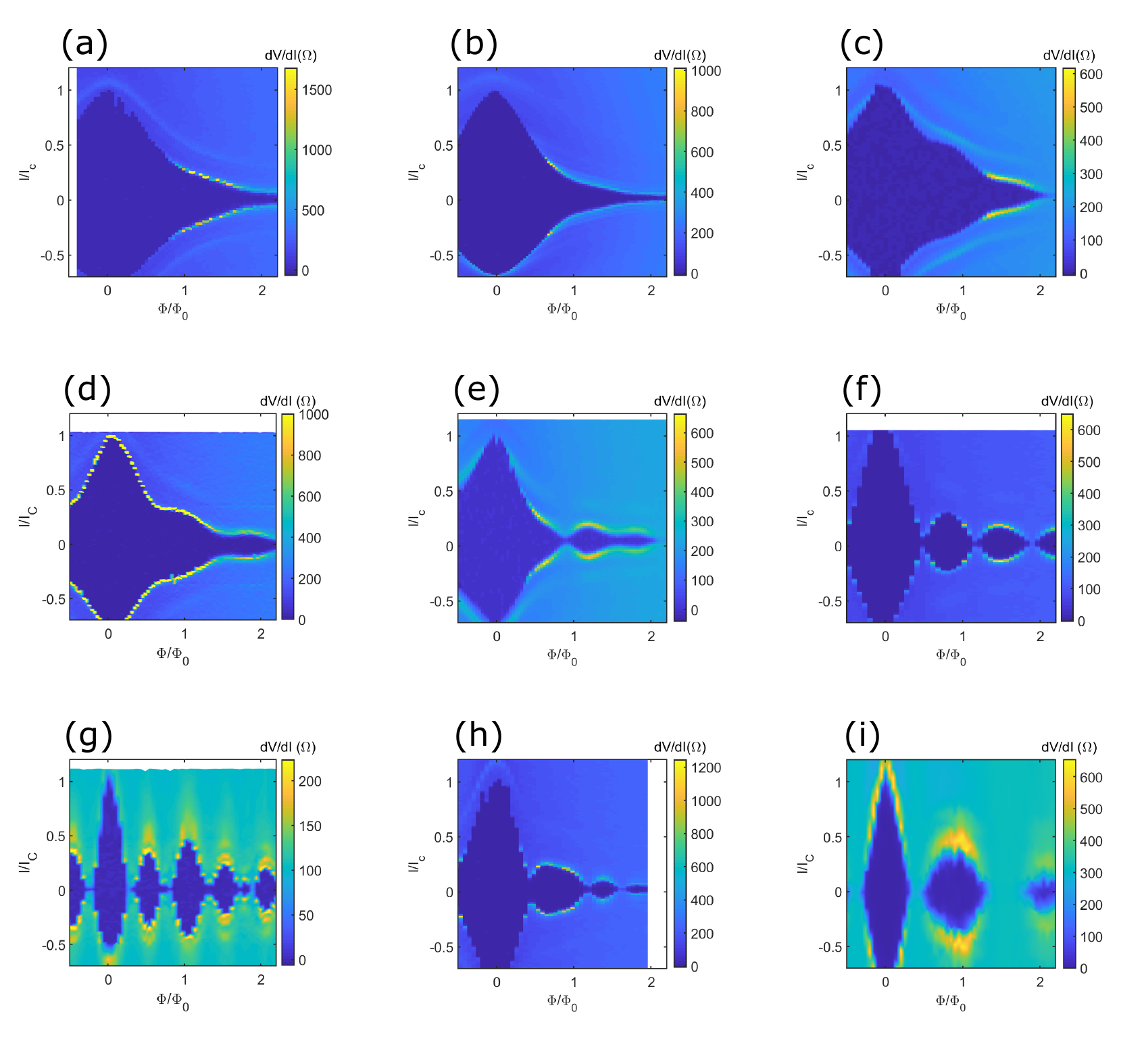}
    \caption{\textbf{Impact of sample transparency on $\boldsymbol{I_C(B)}$-oscillations.} Color maps of the differential resistance $dV/dI$ as a function of the current normalized $I/I_c$ and the magnetic flux $\phi/\phi_0$ for different samples. The color maps are ordered by decreasing transparency of the junctions, with the highest transparency shown in (a) to the lowest in (i). \textbf{a, b,} Samples A and B with high transparency ($D\approx0.70$ and $D\approx0.66$) show no oscillations as a function of the magnetic field. \textbf{c, d, e,} Intermediate transparencies in samples C, D, E ($D\approx0.64$, $D\approx0.63$, and $D\approx0.62$, respectively): the shape of the critical current contour starts to deviate and first nodes and antinodes are observable. \textbf{f, g, h, i,} Samples F - J with the lowest transparencies ($D\approx0.57$ to $D\approx0.43$) show distinct oscillations as a function of applied magnetic field.}
    \label{fig:transmission}
\end{figure*}

\subsection{Gated devices}

Fig.~\ref{fig:gated device}c shows the data of sample J. This device has a critical current $I_C=\SI{136}{\nano \ampere}$ and an average transparency $D=0.43$, while one flux quantum $\phi_0 = h/2e$ corresponds to $B\approx\SI{50}{mT}$. For this sample we also observe $I_C\left(\phi\right)$-oscillations. However, only maxima at $\phi=n\cdot\phi_0$ are visible leading to a $h/2e$-periodicity. For more detailed studies, a top-gate was added to the junction. This allows to investigate the $I_C\left(\phi\right)$-oscillations as function of the gate voltage $V_G$. The structure of a gated device is sketched in Fig.~\ref{fig:gated device}a. An insulator made of $\sim\SI{30}{\nano\metre}$ \ce{SiO_2}, grown by PECVD, and $\sim\SI{100}{\nano\metre}$ \ce{Al_2O_3}, grown by ALD, was deposited above the junction. The top-gate voltage $V_G$ is applied via a metallic \ce{Ti}/\ce{Au}-layer. Fig.~\ref{fig:gated device}b shows the critical current $I_C$ as a function of the top-gate voltage $V_G$. By tuning $V_G$ from $\SIrange{0}{3}{V}$, $I_C$ increases by a factor $\sim1.7$. Fig.~\ref{fig:gated device}d illustrates $dV/dI\left(\phi,\,I\right)$ of sample J for $V_G=\SI{3}{V}$. Additional maxima appear at $\phi=(2n+1)\cdot\phi_0/2$ in contrast to the data at $V_g=0$. Hence, the $h/4e$-periodicity is recovered by increasing $V_G$. This observation emphasizes that the $h/2e$-oscillations are the dominating ones and are observable for any $V_G$. The maxima at $\phi=(2n+1)\cdot\phi_0/2$ cannot be resolved for $V_g=0$ due to the low $I_C$ at these positions. By increasing $V_G$, the number of contributing channels increases and the increased $I_C$ enables to resolve $I_C$ at $\phi=(2n+1)\cdot\phi_0/2$. Compared to sample G, however, $I_C\left(\phi\right)$-oscillations with a period $h/8e$ are not observable, although the transparency of the devices are similar. Sample G was fabricated from a doped wafer. Its electron density, and thus the number of transport channels contributing to the signal, is much higher than in the undoped sample J, even when the latter is gated at high voltages. This suggests that for observing higher harmonics in the $I_C\left(\phi\right)$-oscillations a sufficiently large number of transport channels is necessary.

\subsection{Influence of the transparency}

In addition to differences in geometry, the transparency of the superconductor/nanowire interface is the decisive parameter that differentiates the devices studied. Fig.~\ref{fig:transmission} shows color maps of the differential resistance $dV/dI$ as a function of the normalized current $I/I_C$ for several samples with different transparency. The transparency was  calculated using the $I$-$V$ characteristics, as explained above  and explicitly demonstrated for sample G. Here it should be mentioned that the extracted transparency gives a value averaged over all contributing transport channels. Thus, it can vary locally at the superconductor/nanowire interface. In Fig.~\ref{fig:transmission}, the color maps are ordered by the device transparency, descending from higher to lower values from top left to bottom right, $(a)\to(i)$. Moreover, the labelling of the devices A-J follows the labelling in the figure (a)-(i). Thus, devices A and B have the highest transparencies, $D\approx0.70$ and $D\approx0.66$, among the samples investigated. For these high-transparency devices the critical current $I_C$ monotonously decays with increasing magnetic flux $\phi$. For samples with slightly lower transparency $D\approx0.64$ and $D\approx0.63$, as in samples C and D, the monotonic decrease of the critical current still prevails but an additional shoulder comes out. This shoulder can be considered as a precursor of the supercurrent interference appearing at still lower transparencies. The oscillations start for device E ($D\approx0.62$). Initially, $I_C$ decreases and is almost fully suppressed below $\phi=\phi_0$. Then, $I_C$ increases again and shows a maximum around $\phi=\phi_0$. The oscillations become more pronounced for samples F ($D\approx0.57$), G ($D\approx0.51$), H ($D\approx0.49$), and J ($D\approx0.43$) which have an even lower transparency.

These samples show clear $I_C(B)$-oscillations with periodicities $h/2e$ or $h/4e$. For samples G and J, the maxima appear exactly at positions $\phi=n\cdot\phi_0/2$ and $\phi=n\cdot\phi_0$ respectively, while the positions are slightly shifted for devices E and F, where the observed oscillation periods deviate by about 10 percent of a flux quantum from $h/4e$ and $h/2e$. For sample H, the observed periodicity is approximately 20 percent smaller than what one would expect from geometry. These deviations occur in samples with much wider superconducting contacts (see table \ref{tab:sample-table}, suggesting that the larger contacts might affect the flux distribution in the junction.  

Based on these experimental observations we conclude that the transparency $D$ is the most influential parameter that determines whether $I_C(B)$-oscillations occur or not. The oscillations appear preferentially for samples with low average transparency, while they are fully absent for high transparencies. 


\section{Theory}
\label{sec:theory}

\begin{figure}
    \centering
    \includegraphics[%
      keepaspectratio,%
      width=\columnwidth,%
    ]{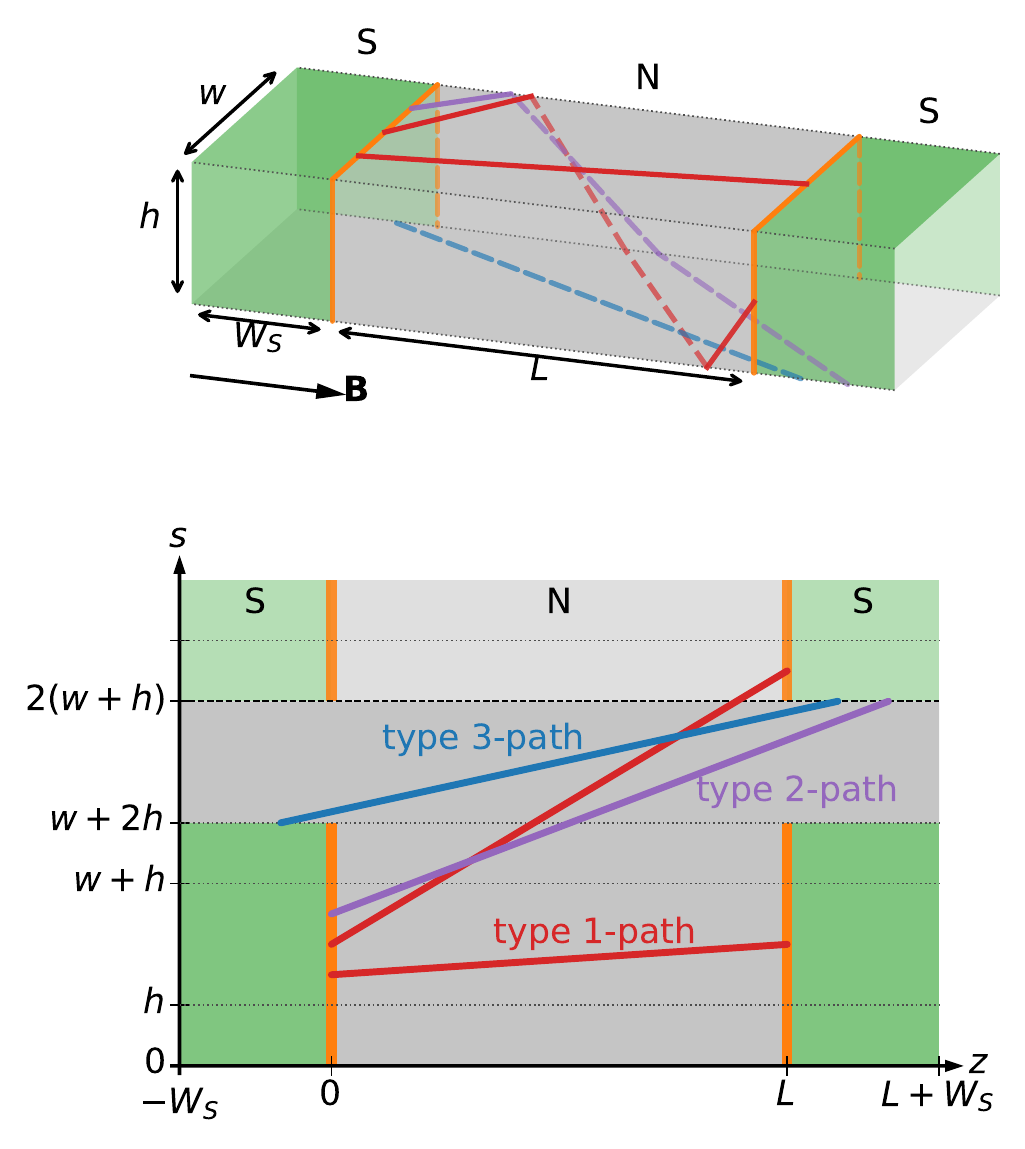}
    \caption{%
        \textbf{Geometry of the system used in the theoretical model.}
        The upper panel shows a 3D sketch of the nanowire Josephson junction
        while the lower panel is a 2D sketch of the rolled out and periodically continued nanowire surface.
        Regions with induced superconductivity are shaded green,
        normal conducting regions gray.
        Superconductivity is not induced around the whole circumference,
        the bottom area is still considered as normal conducting. Additional barriers used in the model are marked as vertical orange lines in the lower panel.
        The different type of retro-reflected paths arising from our semiclassical analysis, Sec.~\ref{sec:semiclassics}, are shown in red, purple and blue, respectively.
    }
    \label{fig:system}
\end{figure}


To proceed we summarize the desiderata and key aspects of the physical problem from a more theory-oriented point of view: 
\begin{itemize}
\item[(i)] there must be sufficiently many open surface channels between the two superconducting electrodes to ensure a fairly high $I_C$;
\item[(ii)] a number of open channels must be sensitive to the flux threading the nanowire cross-section, otherwise no $\phi$-periodicity would show up;
\item[(iii)] imperfect contacts, representing barriers for the transport electrons, suppress contributions from flux-insensitive channels relative to flux-sensitive ones.
\end{itemize}
Based on these premises we first define the model geometry, sketched in Fig.~\ref{fig:system} and introduced in detail below.
An assumption that will turn out to be critical is that the Nb fingers induce superconductivity only close to the contact regions (shaded green areas in Fig.~\ref{fig:system}), \ie the nanowire bottom surface remains normal conducting\footnote{Except for the highest-quality samples, see the discussion in Sec.~\ref{sec:exp_theo_comparison}.}.  We will later demonstrate that modes formed by Andreev retro-reflection (partially)  winding around the circumference of the 3DTI nanowire pick up an Aharonov-Bohm phase and lead to the experimentally observed supercurrent oscillations.
To reach our conclusions we combine semiclassical analytics with tight-binding numerical simulations.  Semiclassics allow us to identify the fundamentals of the transport problems in terms of families of electronic paths which enclose (or do not enclose) a magnetic flux.  This picture is validated by rigorous quantum transport simulations based on a tight-binding implementation of the corresponding Bogoliubov-de Gennes (BdG) Hamiltonian, see below.  
Our analysis shows that the relevant aspects of the problem are geometrical (non-planar surface conduction, winding vs.~straight propagation, nanowire perimeter not fully superconducting), while the Dirac or trivial (quadratic) nature of the carriers seems to play a secondary role.



\subsection{Geometry and Model}
\label{sec:model}

The upper panel of Figure~\ref{fig:system} shows the model geometry of the 3D nanowire JJ and the lower panel its unrolled surface. In the figure,
$w$ and $h$ denote the nanowire width and height,
$L$ the junction length,
and $W_S$ the width of the superconducting contacts. 
We also introduce the perimeter $P = 2 w + 2 h$
and the interfacial boundary $C = w + 2 h$ which describes the length of the perimeter covered by the superconducting contacts.

In our model we include phenomenological delta-like barriers at the interfaces 
between normal and superconducting parts in the transverse direction only, \ie along the perimeter, {\em i.e.},
\begin{equation}
\label{eq:barrier}
    U(z, s) = U_0 \Theta(s) \Theta(w + 2 h - s) [\delta(z) + \delta(z - L)] \,   .
\end{equation}
The barriers are marked in orange in Fig.~\ref{fig:system}.
They account for the fact that the supercurrent oscillations appear in the less transparent
junctions; see Fig.~\ref{fig:transmission}.  Indeed, the barriers turn out to be essential for the observation and understanding of the supercurrent oscillations with a flux-periodicity of $h/4e$.
The reason for their presence lays in the fabrication process itself. 
Foremost, an incomplete removal of the capping layer induces a barrier at the interfaces between Nb and HgTe. 
These complex interface physics are simplified, but essentially captured, by the local delta barriers.

Starting from this geometrical model the JJ system is quantum mechanically described by the Bogoliubov-de Gennes Hamiltonian
\begin{equation}
    \label{eq:BdG-Hamiltonian}
    H = \begin{pmatrix}
        h_e & \Delta e^{i \varphi} \\
        \Delta e^{-i \varphi} & h_h
    \end{pmatrix}
    ,
\end{equation}
where $h_e$ and $h_h$ describe the electron and hole Hamiltonians
and $\Delta$ and $\varphi$ denote the absolute value and phase of the pairing potential.

The topological surface states are described by the Dirac model Hamiltonian \cite{2010-PhysRevLett-Zhang_Vishwanath,2010-PhysRevLett-Bardarson_Brouwer_Moore}
\begin{equation}
    \label{eq:H_Dirac}
    h_{\textit{e/h}} = \pm \hbar v_F \left[
        \qmop{k}_z \sigma_x + \left( \qmop{k}_s \pm \frac{\phi}{\phi_0} \frac{\pi}{P} \right) \sigma_y
    \right] \mp \mu \pm U
    ,
\end{equation}
the upper (lower) sign denoting the electron (hole) Hamiltonian.
Here, $z$ and $s$ are the coordinates along the wire axis and along the perimeter,
and $\qmop{k}_z$ and $\qmop{k}_s$ the respective momentum operators. Furthermore,
$\mu$ is the chemical potential and $U$ denotes the barriers at the NS-interfaces, see also below.

Only the nanowire surface in direct contact with the superconductor, shaded green in Fig.~\ref{fig:system}, is affected by the proximity effect. Its bottom surface, grey in Fig.~\ref{fig:system}, remains normal.
Accordingly, the absolute value of the pairing potential is modelled as follows:
\begin{equation}
    \label{eq:Delta}
    \Delta = \begin{cases}
        \Delta_0 & \text{for $0 \leq s \leq C$ and $-W_S \leq z \leq 0$,} \\
        \Delta_0 & \text{for $0 \leq s \leq C$ and $L \leq z \leq L + W_S$,} \\
        0 & \text{otherwise.}
    \end{cases}
\end{equation}
Furthermore, we assume that the thickness of the \ce{Nb} contacts
is much smaller than the London penetration depth of \ce{Nb}
such that no supercurrent develops around the perimeter
and the magnetic field is not screened.
Thus, the Hamiltonian \eqref{eq:BdG-Hamiltonian} has to be defined in an gauge invariant way.
To ensure this the superconducting phase $\varphi$, defined only in the regions $W_S \leq z \leq 0$ and $L \leq z \leq L + W_S$, satisfies
\begin{equation}
\label{eq:Phase_of_Delta}
    \varphi = \begin{cases}
        - \frac{1}{2} \varphi_0 + 2 \pi \frac{\phi}{\phi_0} \frac{s}{P} & \text{for $-W_S \leq z \leq 0$,} \\
        + \frac{1}{2} \varphi_0 + 2 \pi \frac{\phi}{\phi_0} \frac{s}{P} & \text{for $L \leq z \leq L + W_S$, } 
    \end{cases}
\end{equation}
and the unitary transformation $V(\phi) H(\phi) V^{\dagger}(\phi) = H(0)$ holds for
\begin{equation}
    \label{eq:unitary-transformation}
    V(\phi) = \exp\left(i \pi \frac{\phi}{\phi_0} \frac{s}{P} \tau_z \right)
    .
\end{equation}
The transformation also modifies the boundary condition of the wave function,
\begin{equation}
    \label{eq:boundary-condition}
    (V \Psi)(s + P) = \pm \exp\left(- i \pi \frac{\phi}{\phi_0} \tau_z\right) (V \Psi)(s)
    ,
\end{equation}
necessary for the calculation of the Andreev bound states.

Note that Eq.~(\ref{eq:Phase_of_Delta}) for $\varphi$ 
can also be derived using Ginzburg-Landau theory:
The free energy density is proprotional to the supercurrent $\vec{J}_S = - 2 (e n_S / m) (\hbar \nabla \varphi + 2 e \vec{A})$.
Minimizing $\vec{J}_S$ leads to $\nabla \varphi = - 2 e \vec{A} / \hbar$
\cite{1996-Tinkham-superconductivity,2018-PhysRevB-Wojcik_Nowak,2019-PhysRevB-Winkler_et_al}.



\subsection{Semiclassical analysis}
\label{sec:semiclassics}

\subsubsection{Method}
\label{sec:method}

A semiclassical approach is justified in the limit $k_F L \gg 1$, which is fulfilled in our system, see Sec.~\ref{sec:results}.  We thus follow the procedure from Ref.~\cite{2016-PhysRevB-Ostroukh_et_al}.
First we identify all classical self-retracing trajectories $\Gamma$ that arise from pure retro-reflections at the left and right NS contacts. Such trajectories are thus composed of electron-like and hole-like path segments. Each trajectory $\Gamma$ is then assigned a wave mode bound to a small tube of width $\lambda_F = 2 \pi / k_F$ and contributes a current of $j(\Gamma)$ to the total current.
The total current follows by integrating the contributions $j(\Gamma)$ over all paths $\Gamma$ at the Fermi surface.
Choosing a cut $z = z_{\text{cut}}$ through the normal part,
the paths can be characterized by the $s$ coordinate along this cut
and the axial wave number $k_s$
such that the integral reads \cite{2016-PhysRevB-Ostroukh_et_al}
\begin{align}
    I &= \frac{1}{2 \pi} \int \dd s \int \dd k_s \; j(s, k_s)
    \notag\\
    \label{eq:definition-total-current}
    &= \frac{k_F}{2 \pi} \int \dd s \int \dd \theta \; \cos(\theta) j(s, \theta),
\end{align}
with $\theta$ the path angle with respect to the $z$ direction.

The expression \eqref{eq:definition-total-current} contains a significant simplification: It does not account for specular normal reflection at the NS interfaces, which would modify the definition of the current in terms of paths.
The inclusion of additional paths from such normal reflections
substantially complicates the calculations of $j(\Gamma)$ and $I$ and requires the use of resummation techniques beyond the scope of this work. Moreover, we will establish {\it a posteriori} via quantum mechanical simulations that only perfectly retro-reflected paths are particularly important.
Note also that there is no bending of the paths due to the $B$-field, since the Lorentz force points perpendicular to the nanowire surface. Finally, for simplicity we stick to the short junction limit, $L \ll \xi = \hbar v_F / \Delta_0$, although we expect our findings to qualitatively hold for long junctions as well.

\subsubsection{Classification of the trajectories}
\label{sec:trajectory-classification}

The classical trajectories can be divided into different categories.
First, we can assign a ``crossing number''~$n$ to each path
which counts the crossings through the nonproximitized bottom surface.
Formally, one can define a line cut $s = s_{\text{cut}}$
with $C < s_{\text{cut}} < P$
and count the (directed) crossings through this cut.
We emphasize that this integer $n$ does not correspond to a proper winding number around the perimeter.  It only counts the transverse crossings of the non-proximitized bottom surface.

Second, we can group the paths according to their start and end points,
see Fig.~\ref{fig:system}:
\begin{enumerate}
 \item \textit{Type-1 paths} start and end on the $z = 0$ and $z = L$ NS interfaces;
 \item \textit{Type-2 paths} are ``mixed'' paths,
where start and end points are located on a $z = \textit{const}$ and a $s = \textit{const}$ interface;
 \item \textit{Type-3 paths} comprise paths with start and end points on the $s = 0$ and $s = C$ interfaces.
\end{enumerate}
Type-2 paths can be further subdivided into type-2L and type-2R paths, where type-2L paths start on the $z = 0$ interface and type-2R paths end on the $z = L$ interface. It is important to notice that type-2 and type-3 paths only exist for $n \neq 0$, in other words there are only type-1 paths with $n = 0$.

For given initial coordinates $(s_0, z_0)$ and final coordinates $(s_1, z_1)$, the trajectories are parametrized as
\begin{align}
    \label{eq:trajectory}
    s(t) = s_0 + t \frac{k_s}{k_F}
    \quad \text{and} \quad
    z(t) = z_0 + t \frac{k_z}{k_F}
    ,
\end{align}
where the wave numbers satisfy
\begin{equation}
    \label{eq:relation-ks-kz}
    \frac{k_z}{k_s} = \frac{z_1 - z_0}{s_1 - s_0}
    \quad \text{and} \quad
    k_z^2 + k_s^2 = k_F^2
  \,  .
\end{equation}

\subsubsection{Current contributions}
\label{sec:current-contributions}

To calculate the current contribution $j(\Gamma)$ for each classical trajectory $\Gamma$,
we employ the scattering matrix formalism introduced by Beenakker
for 1D Josephson junctions
 \cite{1991-PhysRevLett-Beenakker}.
In the short junction limit $L \to 0$, one gets for the energies of the the Andreev bound states (ABS) \cite{1991-PhysRevLett-Beenakker,2004-JSupercond-Klapwijk}
\begin{equation}
    \label{eq:abs-finite-transmission}
    E = \pm \Delta_0 \sqrt{1 - \tau \sin^2\left(\tfrac{1}{2} \varphi_0 - \gamma\right)}
    .
\end{equation}
Here, the gauge-invariant phase difference $\varphi_0 - 2 \gamma$ appears \cite{1996-Tinkham-superconductivity}, where
\begin{equation}
    \label{eq:aharonov-bohm-phase}
    \gamma = \frac{e}{\hbar} \int_{\Gamma} \dd \vec{s} \cdot \vec{A}
    = n \pi \frac{\phi}{\phi_0}
    .
\end{equation}
is the Aharonov-Bohm (AB) phase of the classical trajectory.
In Eq.~(\ref{eq:aharonov-bohm-phase}) the parameter $\tau$  depends on the transparency and is different for the different types of paths.
For zero temperature, the current contribution reads \cite{1991-PhysRevLett-Beenakker,2004-JSupercond-Klapwijk}
\begin{equation}
    \label{eq:abs-current-finite-transmission}
    j = \frac{e \Delta_0}{4 \hbar} \frac{\tau \sin(\varphi_0 - 2 \gamma)}{\sqrt{1 - \tau \sin^2(\varphi_0 / 2 - \gamma)}}
    ,
\end{equation}
approaching in the limit $\tau \to 1$
\begin{equation}
    \label{eq:abs-current-with-magnetic-field}
    j = \frac{e \Delta_0}{2 \hbar} \sin\left(\tfrac{1}{2} \varphi_0 - \gamma\right) \sgn\left[\cos\left(\tfrac{1}{2} \varphi_0 - \gamma\right)\right]
    ,
\end{equation}
where $\sgn$ is the sign function.
For the different types $m$ of paths, one obtains different $\tau_m$, namely
\begin{align}
    \label{eq:dirac-tau-1}
    \tau_1 &= \frac{1}{\sin^2(\varphi_N) + X^2 \cos^2(\varphi_N)}
    , \\
    \label{eq:dirac-tau-2-3}
    \tau_2
    &= \frac{1}{1 + Z^2 (1 + Z^2)^{-1} \tan^2(\theta)}
    , \quad \text{and} \quad
    \tau_3 = 1
\end{align}
with the dimensionless barrier strength $Z = U_0 / \hbar v_F$ \cite{1983-PhysRevB-Octavio_et_al}.
The parameters $\varphi_N$ and $X$ are given by
\begin{align}
    \label{eq:dirac-phi-n}
    \varphi_N &= 2 \arctan\left(
        \tfrac{\cos(\theta) + Z \tan(\theta)}{ Z - \sin(\theta) - [1 + Z^2 + Z^2 \tan^2(\theta)]^{1/2} }
    \right)
\intertext{and}
    \label{eq:dirac-X}
    X &= [1 + 2 Z^2 (1 + Z^2)^{-1} \tan^2(\theta)].
\end{align}


\subsection{Numerical simulations}

Besides the semiclassical approach we also employ numerical tight-binding simulations with the Python package Kwant~\cite{2014-Groth_et_al}.
Using the finite difference method, the BdG Hamiltonian Eq.~\eqref{eq:BdG-Hamiltonian} and its components, consisting of nontrivial surface states with a linear dispersion Eq.~\eqref{eq:H_Dirac}, are evaluated on a discrete square grid with lattice constant $a$.
Note that by putting the Dirac Hamiltonian on a lattice, the well known Fermion doubling problem arises \cite{1977-PhysRevD-Susskind,1982-PhysRevD-Stacey,1981-PhysLettB-Nielsen_Ninomiya,1981-NuclPhysB-Nielson_Ninomiya_1,1981-NuclPhysB-Nielson_Ninomiya_2}.
This issue can be circumvented by considering an additional Wilson mass term $H_W=E_{W}a/(4\hbar v_F) (k_z^2+ k_s^2)\sigma_z$
\cite{1977-PhysRevD-Susskind,2016-ApplPhysLett-Habib_Sajjad_Ghosh}, which gaps out the artificial Dirac cones at the borders of the first Brillouin zone.
This term is important to avoid  non-physical inter-valley scattering introduced by the potential barriers $U(z,s)$, Eq.~(\ref{eq:barrier}),  in the JJ. 
Also, regarding these delta barriers, one has to appropriately scale the amplitude for the discrete representation.
This is achieved by fixing $U^{'}_0 = U_0/a$. 

\begin{figure*}
    \centering
    \includegraphics[%
      keepaspectratio,%
      width=2\columnwidth,%
    ]{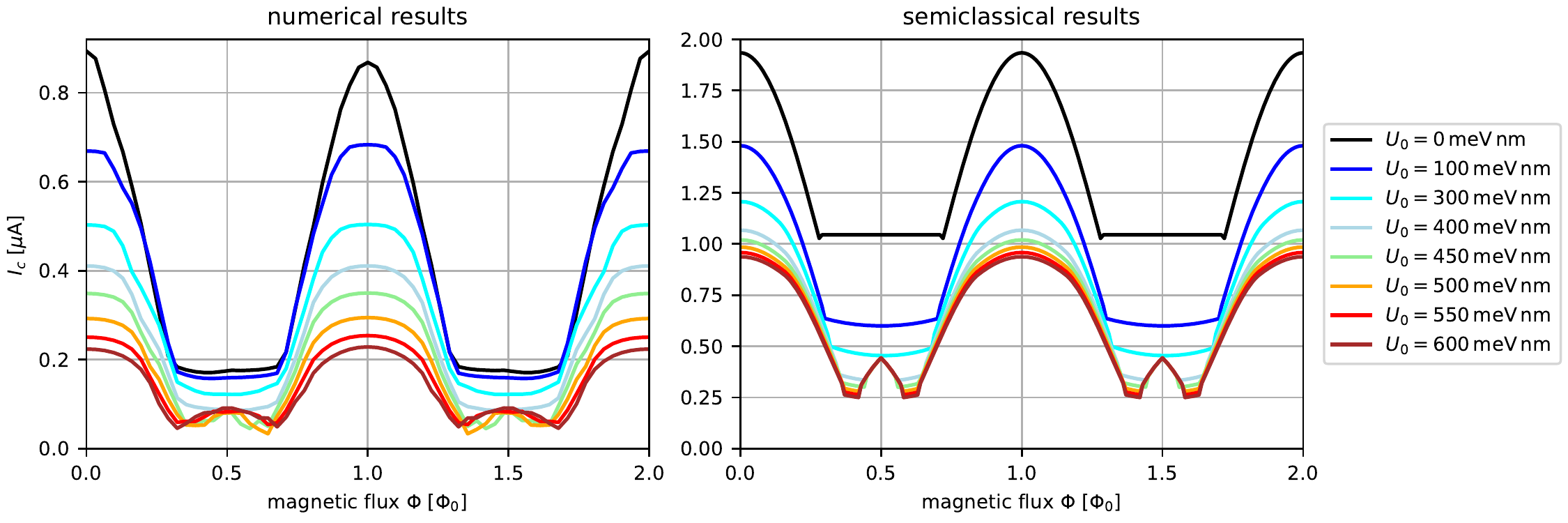}
    \caption{%
      {\bf Critical current for the TI nanowire-based Josephson junction}.
        The results from the semiclassical (left panel) and numerical calculations (right) are shown for four different strengths of the interfacial barrier potential, Eq.~(\ref{eq:barrier}).
        The barrier predominantly suppresses contributions from direct paths which do not cross the bottom surface of the wire,
        such that the peaks at $\phi = h/4e = \phi_0/2$ and $3h/4e=3\phi_0/2$ emerge.
        For larger barrier strengths, those peaks start to appear and become observable in comparison to the peaks at integer multiples of $\phi_0$.
    }
    \label{fig:theory}
\end{figure*}

Connecting the lattice sites with coordinates $(z,s=0)$ and $\left(z,s=P\right)$ by a hopping with phase factor $\exp(i\pi)$ we introduce anti-periodic boundary conditions.
Moreover, the flux through the wire cross section is accounted for by a Peierls substitution with the additional phase factor $\exp(i2\pi\frac{a}{P}\frac{\phi}{2\phi_0} )$.

Finally, superconductivity is introduced as simple onsite s-wave pairing given by Eq.~\eqref{eq:Delta}.
For the numerics we assume semi-infinite leads, {\em i.e.}  $W_s\rightarrow \infty$, because we directly attach translationally invariant superconducting leads to the normal JJ part to keep the numerical cost to a minimum.
Additionally, we consider the local phase modulation of $\Delta$ introduced in Eq.~\eqref{eq:Phase_of_Delta}.

To access the current-phase relation and incorporate all geometrical junction details we compute the supercurrent following Ref.~\cite{FURUSAKI1994214}.
Furthermore we exploit part of the code package provided in a repository of Ref.~\cite{Zuo-et-al-2017}, and adapt it to our implemented tight-binding model. The core of the numerical method is the computation of the current-phase relation via Green's functions.
The supercurrent is given by
\begin{multline}\label{eq:furusaki_I_CPR}
	I_{LR}(\varphi_0,\phi)= 2\frac{ek_{\mathrm{B}}T}{\hbar}\sum_{n=0}^{\infty}
	\sum_{\substack{i\in R \\ j\in L}}\mathrm{Im} \left( H_{ji}G^r_{ij}(i\omega_n)\right. \\
	\left.- H_{ij}G^r_{ji}(i\omega_n) \right),
\end{multline}
where $\omega_n = \frac{k_{\mathrm{B}}T}{\hbar} (2n+1) \pi$ are fermionic Matsubara frequencies.
The labels $i$ and $j$ run over lattice sites in two adjacent transversal lattice rows $R$ and $L$. In Eq.~(\ref{eq:furusaki_I_CPR}) the terms $H_{ij}$ and $G_{ij}$ denote the hopping matrix elements and the off-diagonal elements of Green's function, respectively, connecting those sites.
Furthermore, the phase difference $\varphi_0$ is incorporated into the hopping matrix elements as a phase factor.
This is simply introduced by performing a gauge transformation that shifts the phase difference into a vector potential inside the JJ.
For more details of the methodology we refer the reader to Refs.~\cite{2016-PhysRevB-Ostroukh_et_al,Zuo-et-al-2017}. For a fixed magnetic flux, the critical current is
\begin{equation}
I_c(\phi) = \max_{\varphi_0} |I_{LR}(\varphi_0,\phi)| \, , 
\end{equation}
\textit{i.e.}, the maximum of the corresponding current-phase relation.

\subsection{Semiclassical and numerical results for the critical current}
\label{sec:results}

We are now in a position to combine semiclassics and quantum mechanical simulations to explain the central experimental findings for the critical current reported in Sec.~\ref{sec:exp}.
For the realistic JJ setup discussed in Sec.~\ref{sec:model}  we choose the following parameters to model the SNS-junction geometry, see Fig.~\ref{fig:system}:
$w = \SI{300}{\nano\metre}$, $h = \SI{80}{\nano\metre}$, 
$L = \SI{100}{\nano\metre}$, $W_S = \SI{1000}{\nano\metre}$,
$\hbar v_F = \SI{330}{\milli\electronvolt\nano\metre}$, $\mu = \SI{30}{\milli\electronvolt}$, and $\Delta_0 = \SI{0.8}{\milli\electronvolt}$,
in accordance with Refs.~\cite{2018-PhysRevB-Ziegler_et_al,2021-PhysRevB-Fuchs_et_al,Fischer-et-al-2022}.
The corresponding Fermi wave number is $k_F \approx \SI{0.09}{\per\nano\metre}$, {i.e.}\  the Fermi wavelength  $\lambda_F \sim 70$nm, and $k_F L \approx 10$,. Hence, the semiclassical limit ($k_F L \gg 1$) is well justified.
Since the coherence length reads $\xi \approx \SI{400}{\nano\metre}$,  working in the short junction limit is also justified.
In the semiclassical calculations we include only paths with crossing numbers $n = 0, \pm 1$, since their angle $\theta$ is small and maximizes the $\cos\theta$ factor in the integral~\eqref{eq:definition-total-current}.  Paths with higher crossing number $|n|$ have lower weight, and indeed we checked that including them modifies our results only marginally.  Furthermore higher-crossing paths quickly approach the coherence length cutoff, \ie phase coherence is lost before the electron crosses the junction.  
The value of $\mu$ is chosen to have a high number of open channels while still keeping the numerical simulations in an energetically converged regime.

Numerical and semiclassical results for the critical current are shown and compared to each other in Fig.~\ref{fig:theory}.
On the whole the numerics (left panel) and semiclassics (right panel) show qualitative agreement.
It is convenient to start by looking at the numerics, which show peaks only at integer values of $\phi_0 = h/2e$ in the case of perfect interface transparency, \ie without any barrier ($U_0 = 0$).
We note that in high transparency samples no oscillation was measured at all.  Our theory model predicts no oscillation for fully proximitized systems, which is indeed more likely when the NS junction is good.  In a fully proximitized nanowire there is no phase variation around the perimeter, except in integer multiples of $2\pi$, describing a vortex. Without any kind of accounting for the vector potential in the superconducting phase, the gap and therefor also the critical current will show just an exponential decay.

For increasing barrier strength $U_0$, the interfacial transparencys $\tau_{1,2}$ decrease,
leading to an overall reduction of the critical current.
At the same time, with increasing $U_0$ new maxima emerge and grow at fluxes 
$\phi = h/4e = \phi_0/2$ and $3h/4e=3\phi_0/2$, reaching a peak height of nearly one half the major peaks (for $U_0 = \SI{600}{\milli\electronvolt\nano\metre}$).

The semiclassical results from the right panel of Fig.~\ref{fig:theory} show a corresponding trend: a decreasing critical current with increasing barrier height and the emergence of additional peaks at $\phi = h/4e$ and $3h/4e$. In the semiclassical calculation the dominant peaks arise mainly from the short lead-connecting trajectories marked as type-1 paths with crossing number $n = 1$ in Fig.~\ref{fig:system}. 
Upon increasing the barrier height contributions from such type-1 paths are suppressed relative to those from type-2 and type-3 paths with $n = \pm 1$, since the former involve two barrier reflections while the latter only one, or none at all.
For instance, the current associated with type-3 paths is not influenced by the barrier at all.
The growing relevance of paths with $n = \pm 1$ and no barrier reflection leads to the emergence of the peaks at $h/4e$ and to their increase relative to the peaks at $h/2e$.

To conclude the comparison, semiclassical and numerical results agree on the fundamental aspects: they both predict the emergence of peculiar $h/4e$ peaks for larger barrier strength~$U_0$, the increase of their magnitude relative to the $h/2e$ peaks, and the broadening of all peaks with increasing $U_0$.

A few differences between them, however, remain:
Numerics give a considerably smaller value of the current, 
and the peak current also decreases faster with increasing barrier strength~$U_0$. With regard to this,
first note that the actually induced ''effective'' gap of each of the ABSs as obtained in the numerical calculations is smaller than $\Delta_0$; see App.~\ref{app:ABS_numerics} for a detailed discussion.
To fix this issue in the semiclassical calculation, one would need to introduce an effective gap $\Delta_{\text{eff}} < \Delta_0$ (possibly different for each mode).
Second, as mentioned in Sec.~\ref{sec:method} the semiclassical method neglects contributions from paths with normal specular reflection.
We expect the resulting effects to reduce the current further, as more normal electron reflection reduces the contribution of Andreev reflection.
Furthermore, numerics is not limited to the short junction limit, and in fact fully captures effects of finite length and finite temperature.  For shorter junctions the difference between the semiclassical and numerical current magnitude is indeed smaller, an explicit hint that the short junction assumption of semiclassics loses accuracy for longer systems.

To conclude the theory discussion, the semiclassical appproach is approximate but enables us to interprete the different peculiar peaks in terms of specific (quantized) relevant families of trajectories.
The emergence of the additional peaks related to paths (partially) winding around the nanowire highlights the three-dimensional character of the SNS junction geometry, compared to common planar junctions.

\section{Comparison of experiment and theory}
\label{sec:exp_theo_comparison}

We finally compare the experimentally measured critical currents with the corresponding theoretically calculated results.
Consider first samples with a high average transparency, which can also be modeled in the framework of the effective surface model.  As mentioned in Sec.~\ref{sec:model}, a high quality NS interface should allow for superconducting pairing to be induced across the entire nanowire perimeter. This implies that no phase variation around the perimeter as given by Eq.~\ref{eq:Phase_of_Delta} can develop, and the Andreev bound states states become similar to those of planar Josephson junctions. In such a scenario the magnetic field is simply destroying pair correlations, and the superconducting gap decreases monotonically with increasing field strength.  As a consequence the critical current decays exponentially without any oscillation, as indeed measured in high-transparency samples.

On the contrary, flux periodic supercurrent oscillations are observed in samples with low average transparency.  In Sec.~\ref{sec:model} we argued that the junction transparency might be reduced due to an imperfect removal of a capping layer, which lowers the interface quality between the superconducting Nb and HgTe.  The imperfect interface was modelled both semiclassically and numerically via barriers of varying strength, whose presence suppresses the large current contributions which have no or only a $h/2e$ periodicity. Vice versa, the $h/4e$-periodic current components are not affected and their signatures emerge, providing a clear explanation for the observed behaviour of low-transparency junctions.

\begin{figure}
    \centering
    \includegraphics[width=\columnwidth]{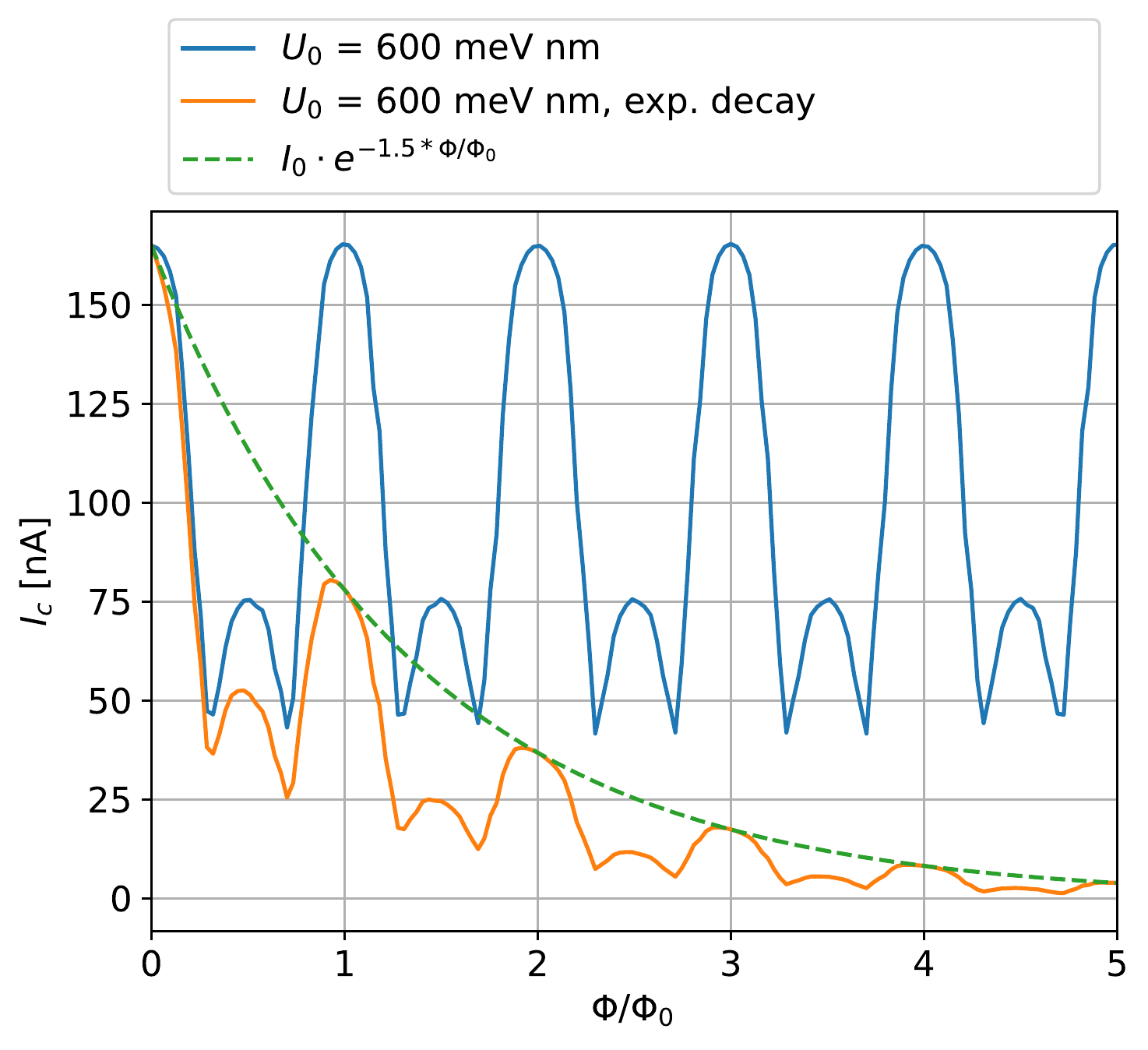}
    \caption{Introducing an exponential envelope function to mimic the pair breaking mechanism of the applied flux leads to a good agreement between theoretical results and the experimental observations. The blue curve corresponds to the originally calculated numerical data, while the orange curve shows the adjusted data.} 
    \label{fig:exp_decay}
\end{figure}
%

Irrespective of the sample quality, all measurements show also a decrease of the current for increasing magnetic field.  This is expected and attributed to the reduction of the induced superconducting gap by the magnetic field \cite{deJuan-et-al-2019}, which weakens pairing correlations. One can phenomenologically account for this behavior by multiplying the theoretical data with an appropriate envelope function, mimicking the weakening of the BdG pairing amplitude $\Delta_0$ of Eq.~(\ref{eq:Delta}).
\footnote{A microscopic description would require a self-consistent treatment of the superconductor in its electromagnetic environment, which is beyond the scope of the present work.}
An example is shown in Fig.~\ref{fig:exp_decay}, where we assumed a simple exponential decay of the pairing potential with respect to the applied flux.
The data was numerically computed with the same system parameters as for Fig.~\ref{fig:theory}, except that the length was increased to $L=200~\mathrm{nm}$ to better match the experimental dimensions. The blue curve is the raw simulation data, while the orange one is adjusted with the phenomenological flux-induced decay. The adjusted critical current exhibits all qualitative features of the experimental curve plotted in Fig.~\ref{fig:transmission}g. In particular the peak at $\phi/\phi_0=1$ is larger than the first half-integer one.

We further remark that also oscillations with a period of $h/8e$ were observed in sample G.
From the semiclassical model such a periodicity is to be expected if paths with crossing number $n = \pm 2$ contribute considerably to the current flow.  This should be possible in the presence of a large overall number of conducting channels, with sufficiently many belonging to the $n=\pm2$ family to make their signature visible -- recall that such paths are identified by a large angle $\theta$, such that the weight of a single path in Eq.~\eqref{eq:definition-total-current} is usually very low.
This agrees with the observation that sample G has indeed the highest number of open transport channels.  Our argument is also in line with the behavior from sample J: A gating potential of $V_G = \SI{3}{\volt}$ has to be applied to the junction, such that the $h/4e$ periodic oscillations can be measured. The gating potential increases the Fermi energy, {\it ergo} the number of open transport channels.  As a consequence the contribution of type-2 and type-3 paths grows and maxima at $\phi=(2n+1)\cdot\phi_0/2$ appear.

\section{Conclusions}
\label{sec:conclusions}

We realized Josephson junctions made of HgTe 3D topological insulator nanowires and demonstrated the fine sensitivity of surface supercurrents to a coaxial magnetic field.  The field does not pierce the topologically protected surface states of the wires, yet Fraunhofer-like critical current patterns develop, notably with unusual non-integer flux periodicity in lower-quality samples.  Our theoretical analysis shows that such peculiar magneto-transport properties are essentially resulting from a series of nontrivial geometrical constraints.  First, contrary to standard Josephson junctions, propagating electronic modes form Andreev bound states uniquely on a non-planar surface enclosing the insulating HgTe bulk. Second, such states may have a purely longitudinal character -- associated with semiclassical paths roughly parallel to the axial direction -- or a partially transverse behavior -- corresponding to  paths winding fully or partially around the wire perimeter -- and are differently affected by the quality of the NS contacts along different directions.  Third, superconductivity is in general not induced across the entire nanowire perimeter, nor is the magnetic field screened by the Nb fingers, which are thinner than the London penetration depth.  As a consequence the partially transverse Andreev bound states pick up an Aharonov-Bohm phase which is not necessarily integer, \ie electrons are not limited to enclosing a fixed number of vortices. This yields the observed peculiar critical current oscillations. 

On the other hand, while the existence of surface states is necessary, spin-momentum locking of topological Dirac states appears to play a minor role. We numerically found similar overall features for surface states obeying an effective Schr\"odinger equation.

For further studies it is certainly desirable and interesting  also to measure the current-phase relation.
Due to the Aharonov-Bohm phase, which is picked up by the Andreev bound states, related signatures could be observable in such measurements and serve as an additional check for the theoretical model.



\begin{acknowledgments}
We thank Denis Kochan, Henry Legg and Philipp R\"u\ss mann for useful discussions. This work was funded by the European Research Council
under the European Union’s Horizon 2020 research and
innovation program (Grant Agreement No.~787515, 253 ProMotion).
We also acknowledge support by the
Deutsche Forschungsgemeinschaft (DFG, German Research Foundation) within Project-ID 314695032 -- SFB 1277 (projects A07, A08, B08)
and through the Elitenetzwerk Bayern Doktorandenkolleg "Topological Insulators". 

\end{acknowledgments}

\appendix
\begin{figure}
    \centering
    \includegraphics[%
      keepaspectratio,%
      width=\columnwidth,%
    ]{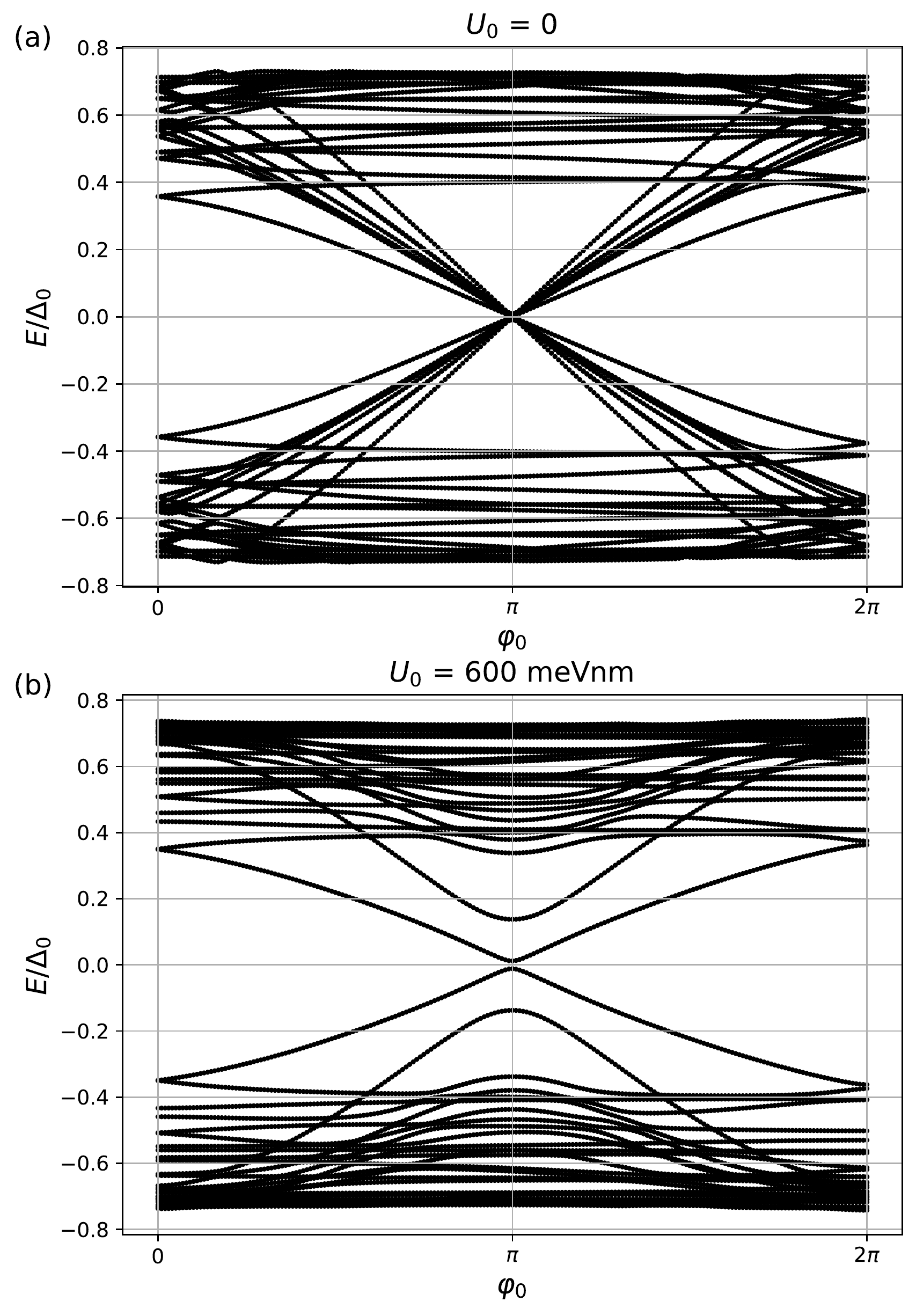}
    \caption{\textbf{Andreev bound state spectrum of a TI Josephson junction which is partially covered by an s-wave superconductor (see Fig.~\ref{fig:system}).} The wire has a width $w=120~\mathrm{nm}$ and a height $h=80~\mathrm{nm}$, while the junctions have a length of $L=200~\mathrm{nm}$. The lattice constant is fixed to $a=5~\mathrm{nm}$ and the chemical potential is set to $\mu=22~\mathrm{meV}$. In (a) the local barrier strength is set to zero, while in (b) the barrier value is set to $U_0 = 600~\mathrm{meVnm}$. For both panels the axial magnetic field is set to zero. The spectrum is computed by diagonalization of a finite tight-binding system. Due to the partial covering the branches in the spectrum have different effective gaps $\Delta_{\mathrm{eff}}$. This explains the difference in the current of the numerical and analytical calculations.}
    \label{fig:ABS_TI_partial_sc_covering}
\end{figure}
\section{Numerical calculation of Andreev bound state spectra for partially covered nanowires}\label{app:ABS_numerics}

The difference in current magnitude of the semiclassical analytical approach and the numerical data can be partially explained by the difference in the spectra of the ABS.
For the analytical approach the ABS spectrum for each mode is assumed to be given by the standard expression Eq.~\ref{eq:abs-finite-transmission}, where the amplitude factor is determined by a constant gap $\Delta_0$.
However, in the numerical calculation, this is not the case.
Due to the partial coverage of the nanowire circumference with the s-wave superconductor, each mode in the surface state spectrum experiences an effective induced gap. 
We can show this by numerically computing the ABS spectrum of such a system.
The eigenenergies of this nanowire Josephson junction can be determined by diagonalization of a finite tight-binding system with long superconducting reservoirs.
The advantage of this method is the natural incorporation of the complex geometry.
The superconducting reservoirs are connected by a periodic boundary hopping, where the superconducting phase difference enters again in a longitudinal hopping in the center of the normal region.
In Figs.~\ref{fig:ABS_TI_partial_sc_covering} (a) and (b) the calculated ABS spectra are plotted for $U_0=0$ and $U_0=600~\mathrm{meVnm}$, respectively.
For simplicity we neglect here the axial magnetic flux and choose a relatively narrow nanowire with width $w=120~\mathrm{nm}$.
The reason for that lies in the reduced number of open subbands, such that the relevant spectrum features are more clearly observable. 
Again we assume that the top, as well as the side surfaces are proximitized by the external s-wave superconductor, while the wire bottom remains normal conducting.
Note that the ABS energies are normalized by $\Delta_0$.
In Fig.~\ref{fig:ABS_TI_partial_sc_covering}~(a) we see that the ABS spectrum remains ungapped due to the missing barriers.
Still, contrary to standard clean Josephson junctions, the ABS branches for different modes are no longer degenerate.
At phase difference zero, where the energies are typically located at the band gap $\Delta_0$, the different branches exhibit very different values.
This indicates that each mode experiences a different effective pairing strength, depending on their angular momentum quantum number.
Also, the values at $\varphi_0=0$ differ quite strongly from $\Delta_0$, which is used in the semiclassical analysis.
Therefore, the difference in current magnitude between numerics and semiclassics can be partially explained by the simplified assumption of a constant superconducting gap in the ABS energies in the latter case.
This holds also true for the case of a non-zero barrier, which is illustrated in Fig.~\ref{fig:ABS_TI_partial_sc_covering}~(b).
The differently induced gaps for each mode are still present, only the spectra become gapped at a phase difference of $\varphi_0=\pi$.

\bibliography{literatur}

\end{document}